\documentclass[11pt, a4paper, pdftex]{article}
\usepackage[utf8]{inputenc}		
\usepackage[T1]{fontenc}

\usepackage{amsmath}
\usepackage{amssymb}
\usepackage{setspace}
\usepackage{latexsym}
\usepackage{fancyhdr}
\usepackage{lastpage}
\usepackage{times}

\usepackage{graphicx}

\usepackage[english]{babel}

\usepackage{authblk}

\graphicspath{{my_figures/}}

\newcommand{\pa}{\paragraph{}}

\def\og{\leavevmode\raise.3ex\hbox{$\scriptscriptstyle\langle\!\langle$~}}
\def\fg{\leavevmode\raise.3ex\hbox{$\scriptscriptstyle\rangle\!\rangle$}}

\begin{document}

\title{Using perceptive subbands analysis to perform audio scenes cartography}
\author[1, 2]{Laurent Millot \thanks{Electronic address: l.millot@ens-louis-lumiere.fr; corresponding author}}
\author[1, 2]{Gérard Pelé}
\author[2]{Mohammed Elliq}
\affil[1]{Institut d'esthétique des arts contemporains - UMR 8592 (CNRS/Université Paris 1/MENRT), Fontenay aux roses, 27 avenue Lombart, 92260, France}
\affil[2]{ENS Louis-Lumière, Noisy-le-Grand, 7 allée du promontoire, 93160, France}

\date{15/03/2005}

\maketitle

	\begin{abstract}
Audio scene cartography for real or simulated stereo recordings is presented. This audio scene analysis is performed doing successively: a perceptive 10-subbands analysis, calculation of temporal laws for relative delays and gains between both channels of each subband using a short-time cons\-tant scene assumption and channels inter-correlation which permit to follow a mobile source in its moves, calculation of global and subbands histograms whose peaks give the incidence information for fixed sources. Audio scenes composed of 2 to 4 fixed sources or  with a fixed source and a mobile one have been already successfully tested. Further extensions and applications will be discussed. Audio illustrations of audio scenes, subband analysis and demonstration of real-time stereo recording simulations will be given.

Paper 6340 presented at the 118th Convention of the Audio Engineering Society, Barcelona, 2005
	\end{abstract}

\section{Introduction}
\pa In this paper we consider the way to analyse audio scene, to determine the number of sources composing it and their potential location. 

\pa Up to now we have studied the case of stereo recordings. Two types of stereo audio scenes have been considered: synthesis of stereo recordings, using a couple of microphones and monophonic pre-recorded sound  files; real recordings of urban ambiances with a real stereophonic couple. 

\pa The first prototype of stereophonic recording simulator has been built by Carr-Brown, Colcy and Delatte \cite{Delatte} during their master thesis supervised by Millot. This prototype has been realized using the Max/MSP programming environment so it is real-time. This simulator is able to record, for the moment in the "free field" assumption, up to four point sources (monophonic sound files) and may constitute an alternative to pan-pot technics used to mix pop, folk or jazz music. For the second prototype of stereophonic recording simulator, part of the current master thesis work of Antoine Valette supervised by Millot,  the microphones models do not take into account the directivity information but should be able to make it appear when it makes sense:  eventually far away from the source. These microphones models, to be presented elsewhere, permit to simulate broadband recordings, as they do not use the monochromatic approach, and to refind a proximity effect. The simulator will be demonstrated during the oral communication. 

\pa The target is the analysis of audio scenes, and as a first step of stereo recordings, in order to determine the incidence of the sources and the potential variations of this incidence with time. For this analysis, a perceptive subbands analysis is performed which provides 10 stereo subbands scenes 
as described in section 2.

\pa In section 3, we explain the main principles of cartography analysis which permits to get the temporal laws for the relative delays ($\Delta t$) and attenuations ($\Delta E$) between left and right channels of the stereophonic scene associated with each subband. From these laws, we show how one may be able to determine the location and number of sources.

\pa Section 4 is devoted to illustrate the cartography process for three different audio scenes : a scene with a static  source and a mobile one; a jazz string quartet ; a live recording in a rail station.

\pa Discussion and perspectives are then adressed in section 5.

\section{Perceptive subbands analysis}
\pa At Louis-Lumière, the research of a sound analysis relevant to sound perception has begun with the master thesis of Boyer \cite{Boyer}. With this first work we studied the possibility to use either the short time Fourier transform or the fast wavelet transform. Even if the wavelet  approach has given better results there was far to much noticeable distorsion introduced in the analysis which prevents us from using listening of the analysis elements as a tool. Furthermore, both Fourier and  wavelet approaches do not rely on human auditory perception.

\pa One may interpret the wavelet analysis technics as filter banks analysis using either downsampling of filtered signals after each new filtering stage (fast wavelet approach) or insertion of  one zero sample between each couple of samples  in the filter impulse responses for each new filtering stage (à trous algorithm). But in both cases, the natural frequential mapping is dyadic which does not correspond to the human auditory one.

\pa Considering the filter bank approach, Millot \cite{Millot} has proposed an analysis method which permits the use of any frequential mapping. At the end of this analysis, one has as many subband sounds as wanted. We must precise that this analysis does not use downsampling because it introduces aliasing, quite noticeable when listening each subband sound. Indeed, one of the constraints of this analysis technic is the possibility to listen any element (subband) of the analysis. At this moment, the filters impulse responses have 8191 samples, which provides rather short transition bands, because it realizes a compromise between filters sharpness and CPU calculation time. The analyser actually runs under Scilab 3.0 on MacOs X or Windows XP but we intend to port it in C or Objective-C by the end of the year.

\pa The sound analyser uses linear phase filters with  finite impulse response which permit to not introduce any phase distorsion and a re-synthesis process by simple mix of the needed subbands: one can choose the subbands he wants to built an approximation of the analyzed sound.

\pa In this paper we consider the extended Leipp frequential mapping \cite{Millot} using the following subbands: 0-50~Hz, 50-200~Hz, 200-400~Hz, 400-800~Hz, 800-1200~Hz, 1200-1800~Hz, 1800-3000~Hz, 3000-6000~Hz, 6000-15000~Hz, 15000-Nyquist frequency (half the sampling rate).  This irregular mapping has been derived by Leipp \cite{Leipp} using musicians and sound engineers and real sounds (music, speech and singing) and may be relevant with the human perception.

\pa Since \cite{Millot}, we have decided to adopt a new  presentation for the frequential information of each channel.  

\pa We still use the Integrated Spectral Density (ISD) which corresponds, for a signal (either a subband signal or the original one), to the sum of the squares of the samples. In fact, with this calculation, we consider the global energy present in the signal.  

\pa But, now, we plot  (in dB) the ratio of the ISD for each channel subband to the ISD for the related original channel. So, one channel plot gives the global relative weight of each subband channel in the generation of original sound channel. 

\pa Figure 1 illustrates this principle for a live recording with an AB ORTF inside the "Gare du Nord" rail station. This figure permits to point out the importance of the first subband (0-50~Hz): very low frequencies are present in everyday life and, according to us,  are the main feature in sound generation for instance in the case of wind instruments.

\begin{figure}[tb!]
\begin{center}
\includegraphics[width=7.5cm]{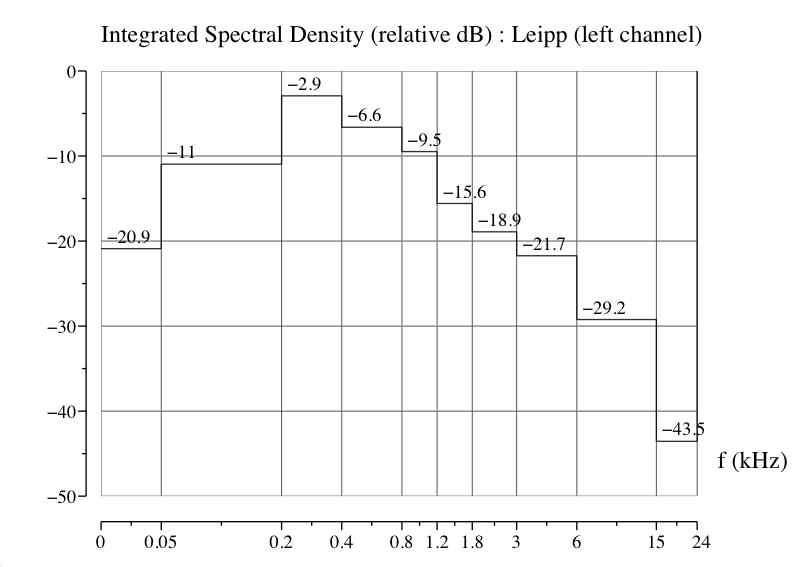}
\includegraphics[width=7.5cm]{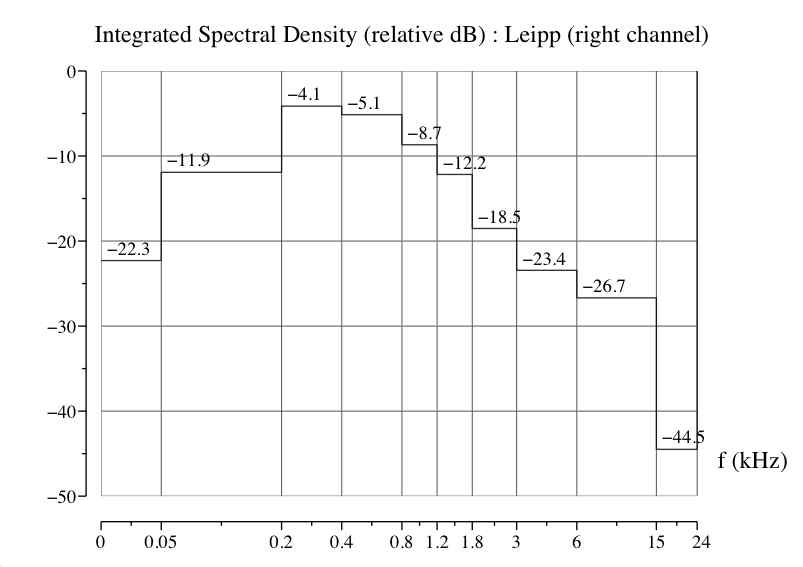} 
\caption[1]{Plots of the left and right ISD  for a real audio scene recorded in a rail station.}
\end{center}
\end{figure}

\section{Cartography principles}
\pa As said before the cartography process consists in determining the number of sources and potential incidence of each sound source composing the audio scene.

\pa The principles, explained here, should be extended to other recording situations than the stereophonic couples, for instance to multichannel recording devices.

\pa To determine the audio scene composition, we use the analysis of the stereophonic recording as 10 limited band stereophonic signals. 

\pa We consider both microphones and each potential source as points. With this assumption, each source excites both microphones with delay and attenuation. For instance, if the source signal first arrives on the left microphone and excites it with a signal $s(t)$, the right microphone will be excited by the signal $A.s(t-\tau)$ where $A$ is a relative attenuation coefficient and $\tau$ a relative delay. $A$ and $\tau$ would depend on the propagation conditions and on composition of the stereophonic couple (nature of the microphones composing the couple). In fact, one can see that we consider  spherical propagation of a broadband wave from a point source to a captation point: the attenuation coefficient may then be related to the distance relying the source and the captation point; the interchannel delay will corresponds to the difference in arrival times at the left and right microphones.

\pa With such an hypothesis we follow the approach adopted by Avendano and Jot \cite{Avendano} and Baskind \cite{Baskind}. But we do not use the short time Fourier transforms as calculation tool and, furthermore, as we use broadband decomposition of the signals, we do not work on sinusoids but on sound signals, which can be listened. 

\pa We can precise that to design the microphone simulator we did not retain the classical monochromatic models, using directivity, but broadband models relying on the propagation of any spherical wave emission from point sources. The microphones models are then constituted using the mix of up to three filtered broadband monopoles.

\pa As we consider that the left and right channels of each subbands are linked by interchannel delay and attenuation we calculate the correlation between both channels for each subbands. This calculation is performed assuming that the audio scene does not significantly change during a short period of time. A 50~ms static scene assumption has given rather interesting results but we have also tested other durations (mainly 10~ms).

\pa After this calculation process, we have 10 temporal laws of interchannel delays ($\Delta t$) and 10 other temporal laws of interchannel attenuation ($\Delta E$): one delay law and one attenuation law for each subband.

\pa To determine the delay and attenuation associated to each potential source, we then calculate interchannel histograms for the delays and attenuations laws to study if some sources can point out. Depending on the situation it can also be interesting to consider the global histograms for interchannel delays and attenuations calculated as the sum of all subband histograms.

\pa For the moment, we are often able to determine  ($\Delta t$, $\Delta E$) couples associated with sources  for synthesized and real audio scenes. With the first version of the microphones models, still using the directivity information \cite{Delatte2}, it was possible to determine the potential incidence directions given a ($\Delta t$, $\Delta E$) couple. But, with the new models of microphones, which do not retain the directivity information, we have been  still working on the problem to determine the source location given a ($\Delta t$, $\Delta E$) couple. Some encouraging results have been already obtained for pure $\Delta t$ couples of microphones.

\pa Using the temporal laws for interchannel delays and attenuations, one can imagine to extract the motion laws of some sources and use them to spatialize other sources.  It should then become possible to use interesting new moves for sound sources, moves which would be hard or impossible to create with classical sound engineering tools. 
 
\pa In the following section, the whole process is applied to three different audio scenes which can give an idea of the potential of the proposed method.

\pa All the scenes presented in this paper use a simulated or real AB ORTF recording because this stereophonic system gives access to $\Delta t$ and $\Delta E$ information but we also  work on the other stereophonic couples.

\section{Applications}
\subsection{Two-sources scene}
\pa The first audio scene studied is composed of a static bass located at 1~m and +45$^{o}$  from the center of the stereophonic couple. For the AB ORTF couple, such location gives an interchannel delay of $\Delta t=0.35$~ms and an interchannel attenuation of $\Delta E=9.2$~dB: the source excites first and stronger the left microphone.

\pa The other source is an organ which describes a circle of radius 1~m around the stereophonic couple whose center is the center of the stereophonic couple.

\pa Figure 2 illustrates the ISD repartition of global energy for each subband and one may  notice that the low frequencies are quite important which is normal because both sources have quite significant low frequency content.

\begin{figure}[tb!]
\begin{center}
\includegraphics[width=7.5cm]{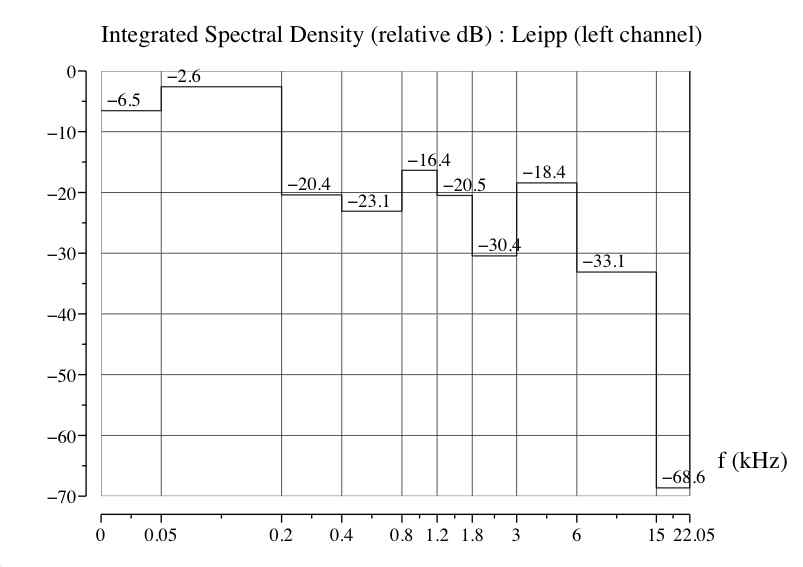}
\includegraphics[width=7.5cm]{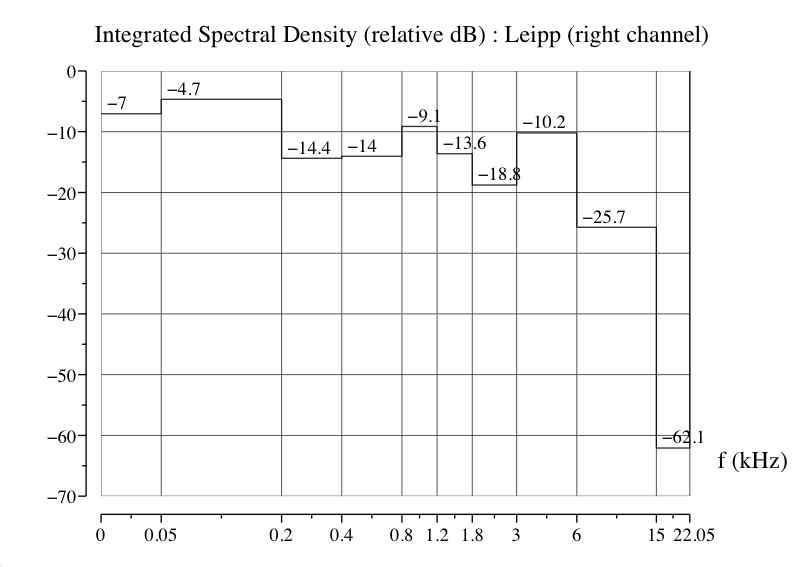} 
\caption[2]{Left and right ISD  for a synthetic audio scene: static bass, organ moving on a circle around the stereophonic couple.}
\end{center}
\end{figure} 

\pa Figure 3 gives the global histograms, respective sums of the subbands interchannel delays histograms and of the subbands interchannel attenuation histograms. One can notice that the information for the static source are obviously present on the histograms. This seems normal because a static source playing a significant role in the audio scene,  will always give the same interchannel delay and attenuation when it plays alone.

\begin{figure}[tb!]
\begin{center}
\includegraphics[width=7.5cm]{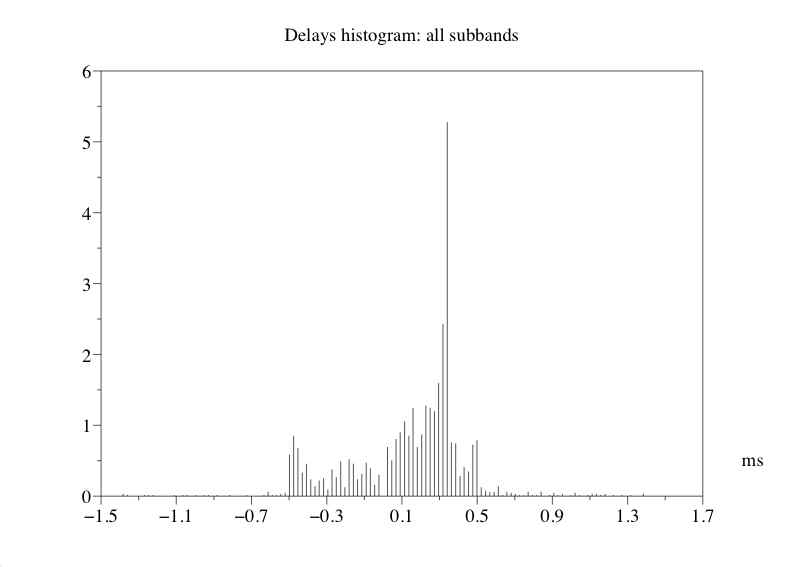}
\includegraphics[width=7.5cm]{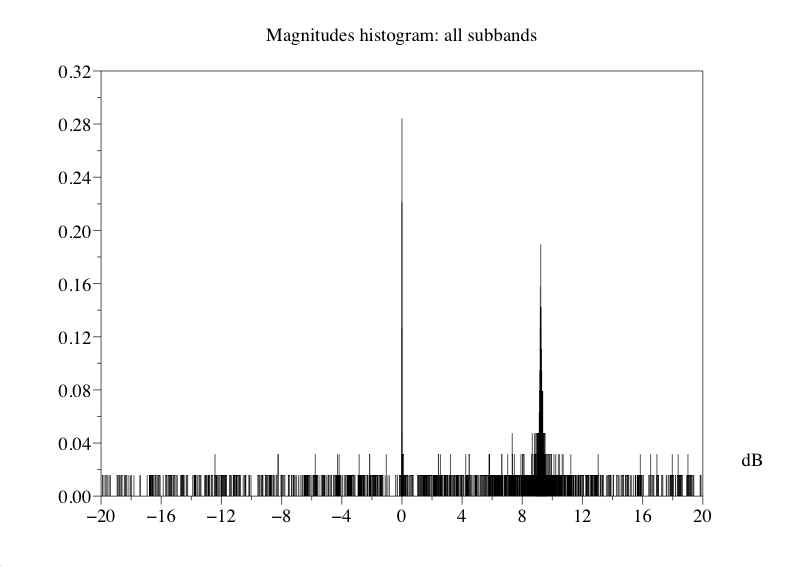} 
\caption[3]{Global histograms for the two-sources synthetic audio scene: relative interchannel delays (left) and attenuations (right).}
\end{center}
\end{figure} 

\pa On figure 3, there is no real information for the moving organ source except a strong peak at 0~dB in the interchannel attenuation histogram (right subplot). This seems also normal because a moving source would give time-varying interchannel delays and attenuations. This can explain the relative extent of both histograms. We think that the peak at 0~dB present in the interchannel attenuation histogram could be related to several locations over the circle for which both microphones are excited at the same time and could also correspond to a zero mean attenuation  for some of the  analyzed "static scenes".

\pa As seen we have found the information for the static source but the information for the moving source is still unknown. And, as the source is moving, this information should be time-varying.

\pa In fact, we can extract the information for the moving source simply by considering a subband where this moving source is alone. This the case for subband 5 (800-1200~Hz) high enough for the bass to be not significantly present. 

\pa Figure 4 present temporal laws for the time-varying interchannel delays and attenuations. The fact that the bass is not significantly present is confirmed by the absence of any offset for both laws: when alone and if present, the bass should have given a 0.35~ms delay and a 9.2~dB attenuation which is not the case here.

\begin{figure}[tb!]
\begin{center}
\includegraphics[width=7.5cm]{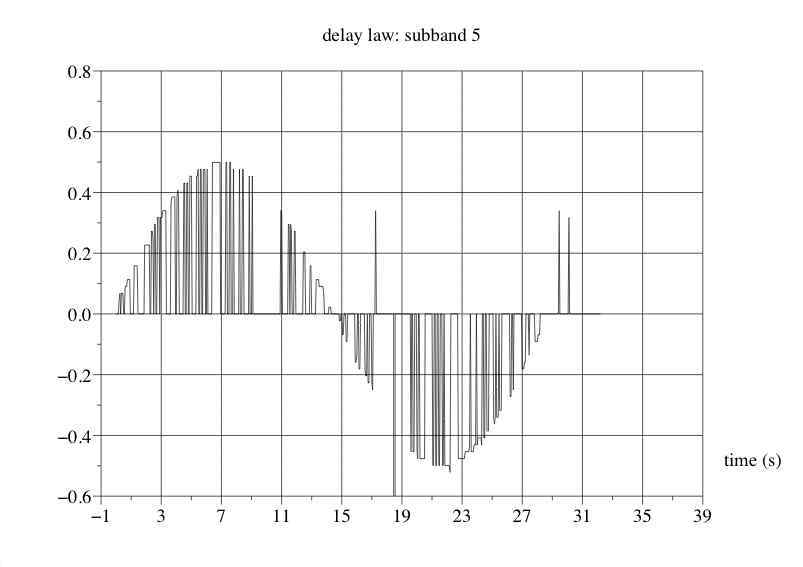}
\includegraphics[width=7.5cm]{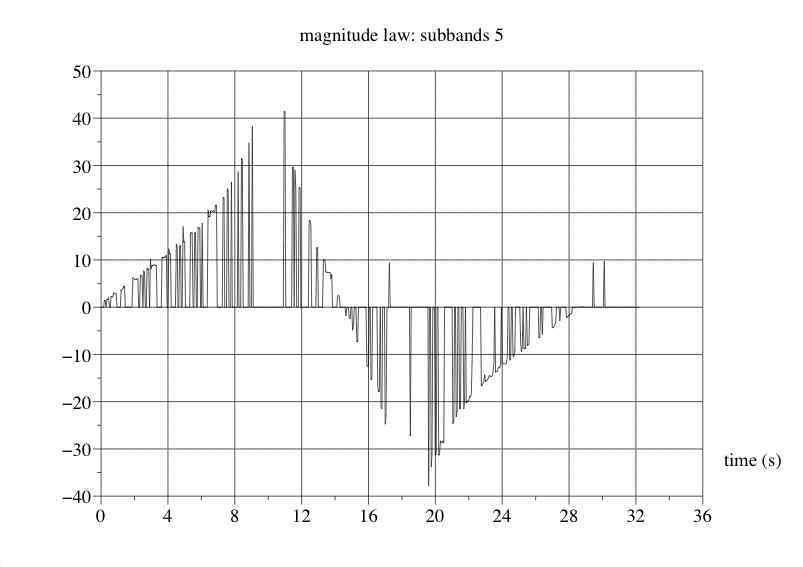} 
\caption[4]{Temporal laws for the subband 5 (800-1200 Hz) in the case of the two-sources synthetic audio scene: relative interchannel delays (left) and attenuations (right).}
\end{center}
\end{figure}

\pa With plots of figure 4 we are potentially able to follow the organ during its motion. To use these laws to spatialize another source it seems necessary to derive continuous versions of the delays and attenuations laws. If this is possible, one can spatialize sound sources following motions which are quite  hard or impossible to manage with classical panning tools.

\pa To derive continuous versions of the laws, the first idea can be  low-pass filtering of the laws but as the motions are quite slow, the cut-off frequency needs to be equal to a few Hertz. Even with longer impulse responses (up to 65535 samples) the filtered version off the laws is unusable: the transitions are less violent but there are still some zero portions, associated with a muted source.

\pa Another solution could be using backward and forward linear predictive analysis techniques in order to re-interpolate the missing "muted" portions. This solution has not been tested for the moment.

\pa In fact, we think a third possibility could be more interesting and perhaps more effective in a production process. 

\pa This third solution consists in building a library of interesting motions around each kind of recording setup and trying to determine if the unknown motion is or not a scaled version of a motion stored in the library. The identification process would use a neural network analyser with training phases (identification of the library elements) composed for instance of a multilayer perceptron. We intend to test this solution in next steps.

\pa To determine which are the meaningful trajectories to include in the library, simulation of stereophonic recordings of real sounds seems an helpful tool. To get  continuous delays and attenuation laws to store, moving some noisy sound sources around the stereophonic couple can be an option. Another option could be, after the determination of the needed motion, the calculation of these laws for each point of the trajectory using the microphones models: calculation of the $\Delta t$ and $\Delta E$ laws for the trajectory points. This method could be rather effective as it does not take into account the velocity parameter which may makes it rather generic: in the library, the meaningful motions would be stored as lookup tables describing the evolution of ($\Delta t$,$\Delta E$) couples all over the motion without any assumption  on the sense of the trajectory or on the speed needed to go from one point of the trajectory to the following one. This is the reason why we would first test this method rather than the linear predictive one.

\subsection{Four-sources scene}
\pa The second audio scene studied in this paper is a string jazz quartet composed of:
\begin{itemize}
\item a double-bass located at 1~m and -10$^{o}$ which gives a $\Delta t$ of -0.086~ms and a $\Delta E$ of -1.99~dB;
\item a banjo located at 1.5~m and -35$^{o}$ which gives a $\Delta t$ of -0.283~ms and a $\Delta E$ of -6.64~dB;
\item a first guitar located at 1~m and 25$^{o}$ which gives a $\Delta t$ of 0.208~ms and a $\Delta E$ of 5.01~dB;
\item a lead guitar located at 50~cm and 40$^{o}$ which gives a $\Delta t$ of 0.315~ms and a $\Delta E$ of 9.5~dB.
\end{itemize}

\pa As this four sources have similar transients and rather similar spectra, this audio scene will be quite difficult to analyse.

\pa Figure 5 presents the ISD for both channels which give the relative weight of each subband in the whole audio sequence. From these two plots, we constat that the subband 2 (50-200~Hz) is the main contributor for the energy.

\begin{figure}[tb!]
\begin{center}
\includegraphics[width=7.5cm]{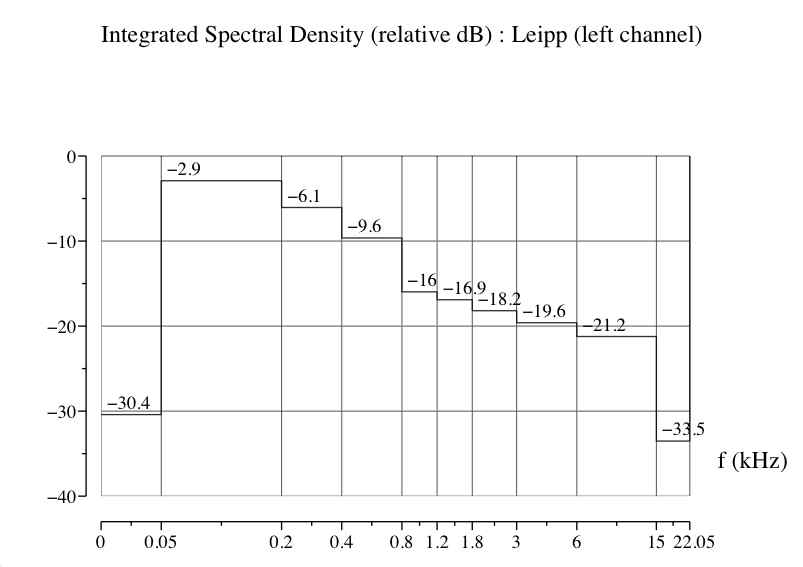}
\includegraphics[width=7.5cm]{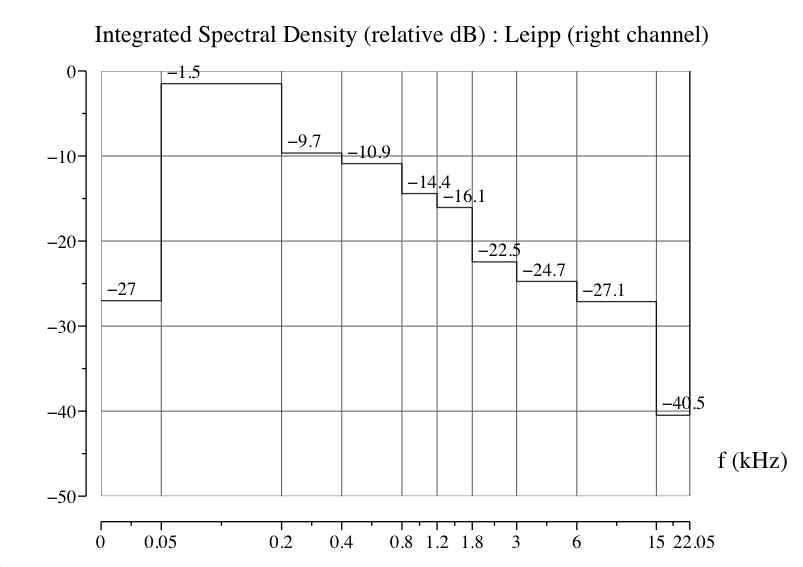} 
\caption[5]{Left and right ISD  for a synthetic audio scene: static bass, solo guitar, guitar and banjo .}
\end{center}
\end{figure} 

\pa Figure 6 gives the global histograms for the interchannel delays and attenuations. As we know how the audio scene was constituted we can notice that some information are potentially already available:
\begin{itemize}
\item peaks for the lead guitar are obviously present in both histograms;
\item a clear peak is present in the delays histograms for the second guitar and an a small hump indicates the presence of this guitar in the attenuations histogram;
\item for the double-bass, there is a clear peak in the attenuations histogram located at the right position but  it is quite difficult to notice the evidence of the double-bass source in the delays histogram because the related hump is tiny;
\item for the banjo, if we find a clear peak in the delays histogram, the attenuation peak  is not so obvious,  moreover in the case where the audio scene composition is unknown.
\end{itemize}

\begin{figure}[tb!]
\begin{center}
\includegraphics[width=7.5cm]{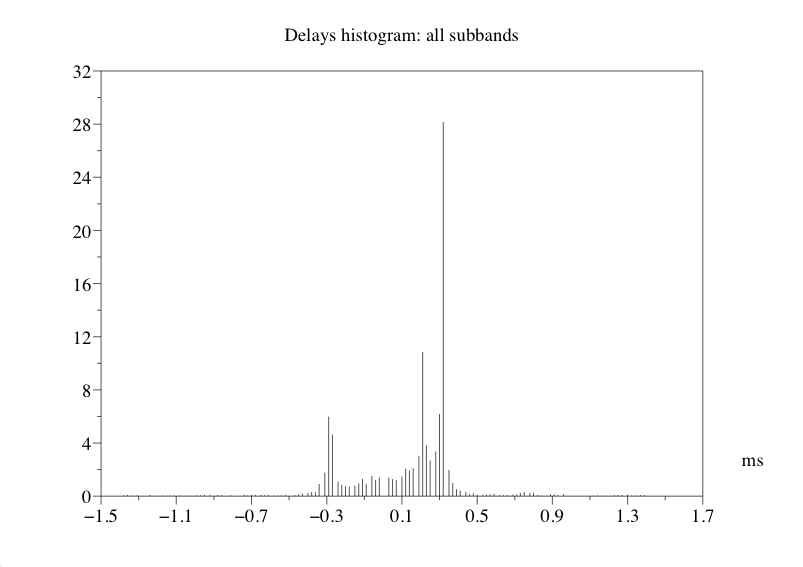}
\includegraphics[width=7.5cm]{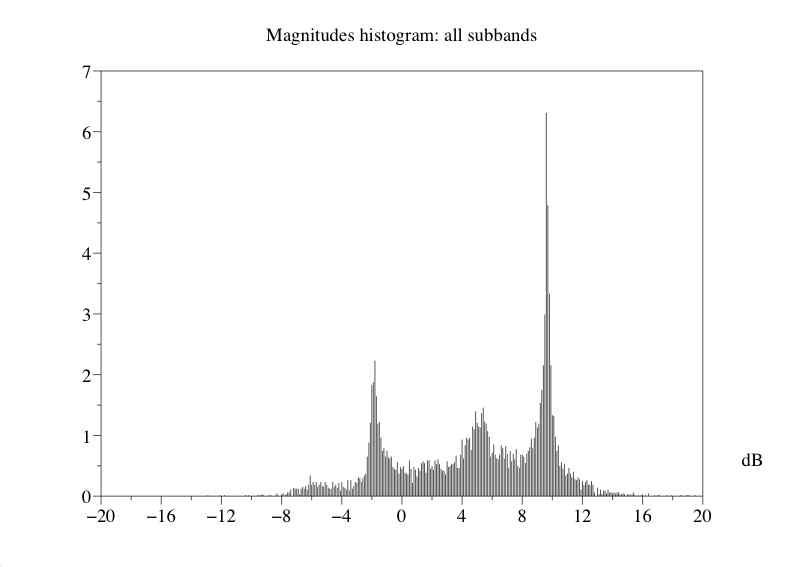} 
\caption[6]{Global histograms for the four-sources synthetic audio scene: relative interchannel delays (left) and attenuations (right).}
\end{center}
\end{figure} 

\pa So, with the study of the global delays and attenuations histograms, we are able to propose that at least three sources are composing the audio scene. The study of some subband histograms can help us, first,  to associate the delays and attenuations peaks, second to precise the number and location of the sources.

\begin{figure}[tb!]
\begin{center}
\includegraphics[width=7.5cm]{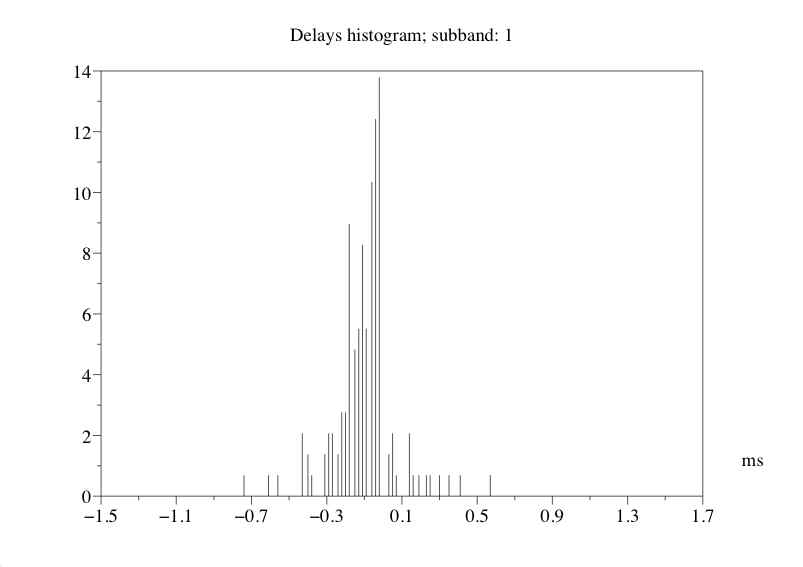}
\includegraphics[width=7.5cm]{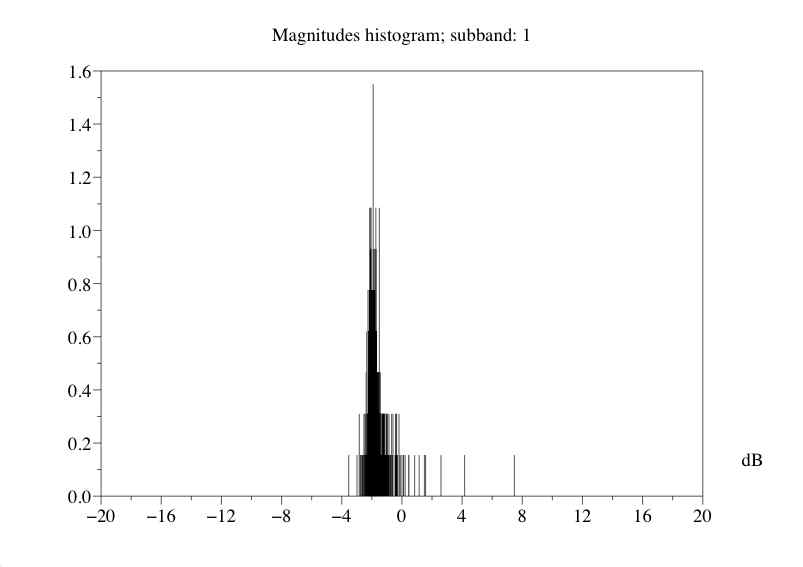} 
\caption[7]{Subband histograms for the subband 1 (0-50 Hz) in the case of the four-sources synthetic audio scene: relative interchannel delays (left) and attenuations (right).}
\end{center}
\end{figure} 

\pa The first subband to consider is the subband 1 (0-50~Hz) where the only source to be present (when listening) is  the double-bass. Figure 7 presents the delays and attenuation histograms for this subband and one can notice that the presence of only one source which permits to associate the $\Delta t$ and $\Delta E$ values which emerge from both plots.   

\begin{figure}[tb!]
\begin{center}
\includegraphics[width=7.5cm]{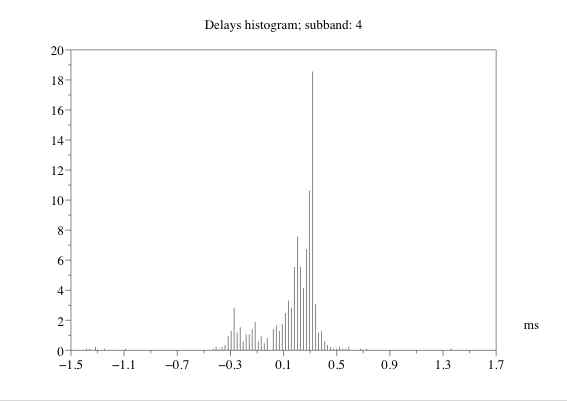}
\includegraphics[width=7.5cm]{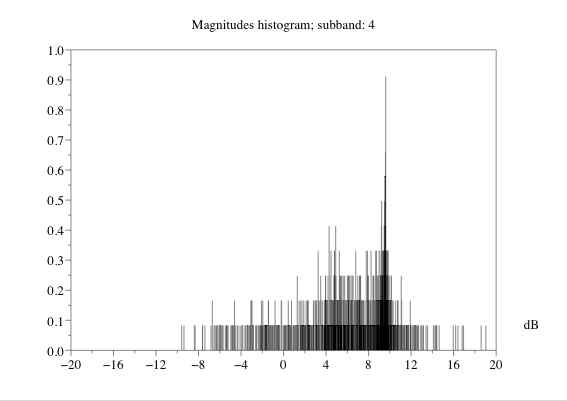} 
\caption[8]{Subband histograms for the subband 4 (400-800 Hz) in the case of the four-sources synthetic audio scene: relative interchannel delays (left) and attenuations (right).}
\end{center}
\end{figure} 

\pa Considering the case of the subband 4 (400-800~Hz), illustrated by figure 8, we can see that all the information to find the solo guitar are available as two clear peaks, one in the delays histogram and the other in the attenuations histogram, arise from the plots. At this point, the information for the solo guitar and for the double-bass location are identified. 

\begin{figure}[tb!]
\begin{center}
\includegraphics[width=7.5cm]{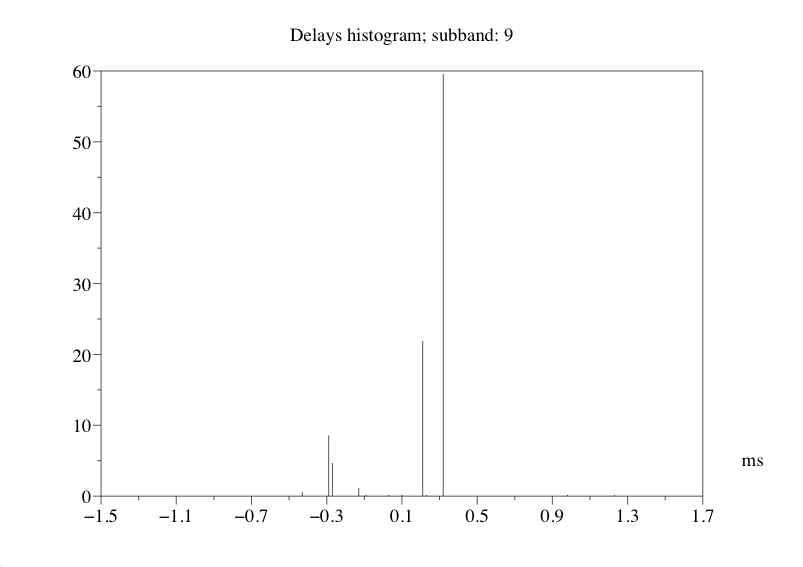}
\includegraphics[width=7.5cm]{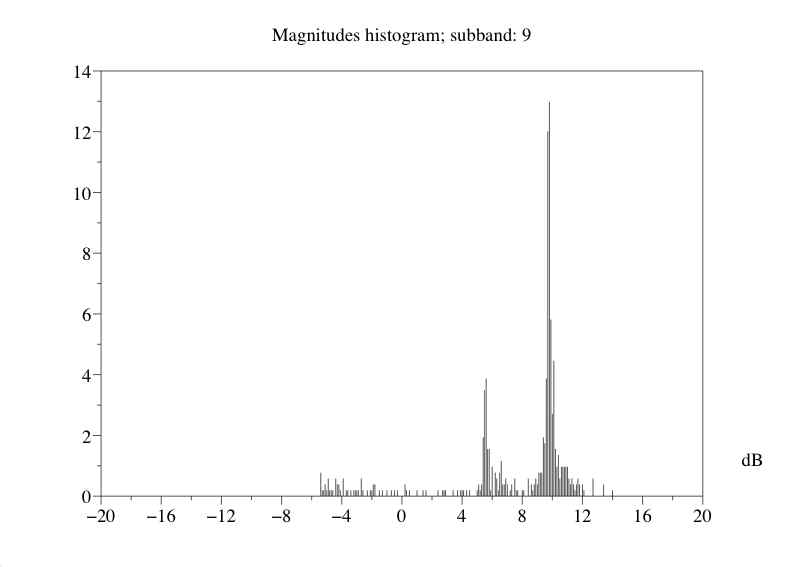} 
\caption[9]{Subband histograms for the subband 9 (6-15 kHz) in the case of the four-sources synthetic audio scene: relative interchannel delays (left) and attenuations (right).}
\end{center}
\end{figure} 

\pa To find information about the first guitar, which plays the "pump", we can consider the histograms associated with subband 9 (6-15~kHz), plotted in Figure 9. Two clear peaks are systematically present in both histograms but as we already have identified the peaks for the solo guitar we can directly derive  the information for the second guitar. Up to this point we have found three sources but, as we found another interesting peak in the global delays histogram, we know that we need to find at least another source.

\begin{figure}[tb!]
\begin{center}
\includegraphics[width=7.5cm]{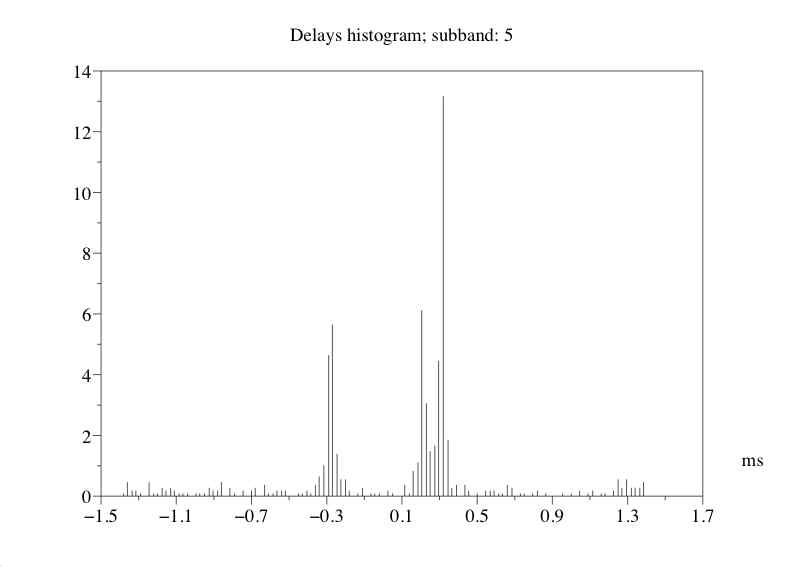}
\includegraphics[width=7.5cm]{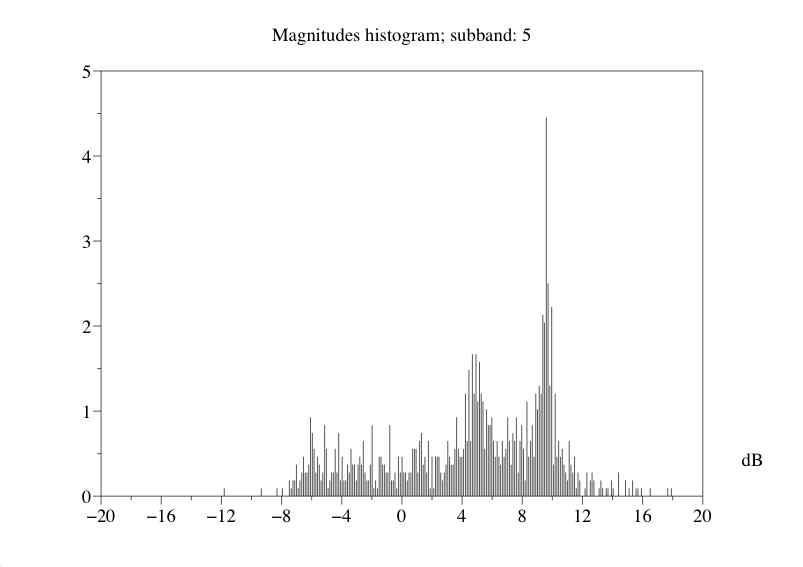} 
\caption[10]{Subband histograms for the subband 5 (800-1200 Hz) in the case of the four-sources synthetic audio scene: relative interchannel delays (left) and attenuations (right).}
\end{center}
\end{figure} 

\pa The investigation can then continue by the study of subband 5 (800-1200~Hz) information, presented in figure 10. The peak in the delays histogram is quite clear for the fourth source, the banjo, and even clearer than in the global delays histogram. But, for the attenuation the situation is disappointing as we are not able to point out the attenuation associated with the banjo. The best emergence of the attenuation peak is found in the global attenuation histogram.

\pa So the fourth source is quite difficult to localise without further information which may be explained by the fact that the banjo is the farest source in the stereophonic recording.

\pa To find out this fourth source, a first solution could consist in re-analyzing the subband where this source is the most powerful but nothing can guarantee that this process may be successful if the sources are too much close in temporal and frequential content.  

\pa Another solution could be to use a parametric analysis of the subband where this source is the most powerful but, for this solution, it will be necessary to use some well designed signals models. As we consider a piece of music, using a sinusoid basis for the signal models will require to introduce a partial following algorithm to be able to characterize the source. This enhancement of the cartography process has not yet been tested but using a subbands analysis would help to use quite simple signal models.

\subsection{Station scene}
\pa The third audio scene presented to illustrate the cartography process is a live recording at the "Gare du Nord" rail station during an announcement of a train information. Due to the acoustical environment, the audio scene is highly reverberated which may complicate the cartography process because of the numerous echoes. 

\pa Yet, this recording is presented to demonstrate that even for such rich audio scene, the cartography process can supply us some interesting information. Figure 11 gives the ISD for both channels.
 
\begin{figure}[tb!]
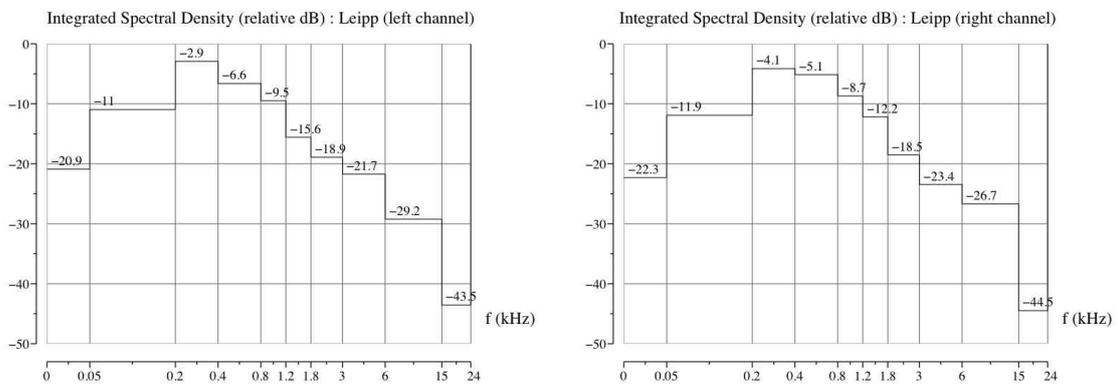

\begin{center}
\includegraphics[width=7.5cm]{Par_6_GdN2_left.jpg}
\includegraphics[width=7.5cm]{Par_6_GdN2_right.jpg} 
\caption[11]{Left and right ISD  for a rail station real audio scene.}
\end{center}
\end{figure}

\begin{figure}[tb!]
\begin{center}
\includegraphics[width=7.5cm]{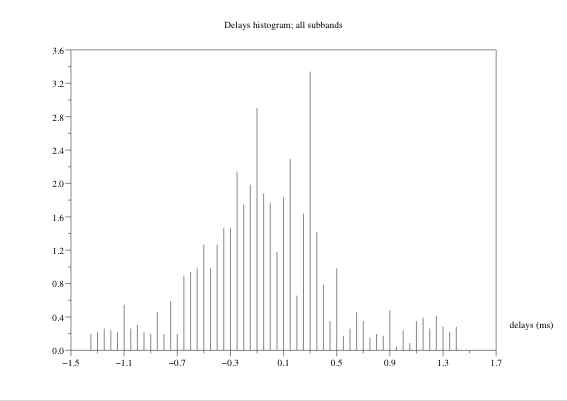}
\includegraphics[width=7.5cm]{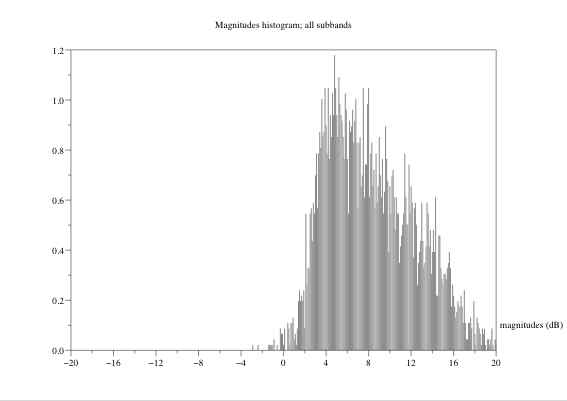} 
\caption[12]{Global histograms for the rail station real audio scene: relative interchannel delays (left) and attenuations (right).}
\end{center}
\end{figure} 

\pa Global delays and attenuations histograms are given in Figure 12 which permit to constate that there are potentially lots of sound sources present in this audio scene, and that perhaps some of them can not be easily represented as point sources. Nevertheless, some peaks may point out from the delays histogram which may indicate that some sources can be identified.

\pa In the following, we consider three particular subbands which clearly permit to identify three kinds of sources. 

\pa Analysis of subband 8, as presented by figure 13, clearly indicates  the presence of a point source. Listening of this subband let us know that this source corresponds to the announcement.

\begin{figure}[tb!]
\begin{center}
\includegraphics[width=7.5cm]{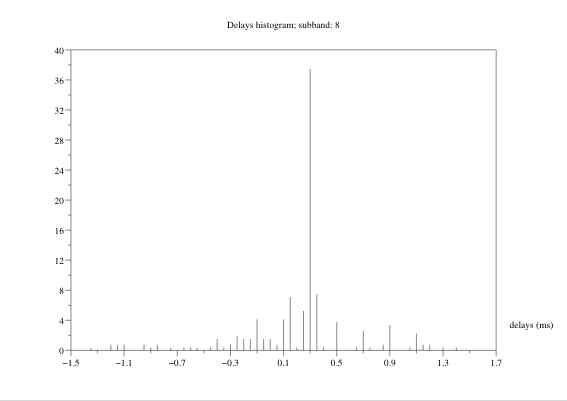} 
\includegraphics[width=7.5cm]{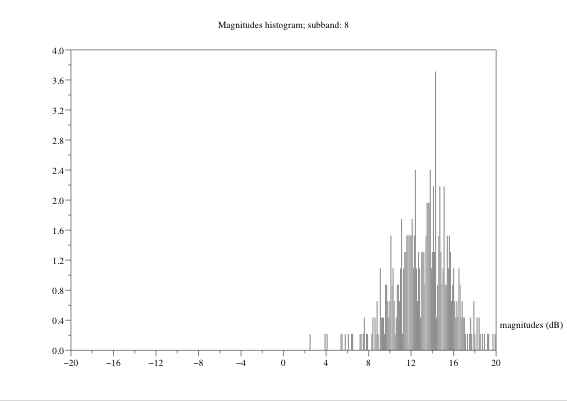} 
\caption[13]{Subband histograms for the subband 8 (3-6 kHz) in the case of the rail station real audio scene: relative interchannel delays (left) and attenuations (right).}
\end{center}
\end{figure} 

\pa Study of the subband 4, illustrated on figure 14, affords the identification of another point source with a different location. Listening of this subband permits to notice that this source is another version of the announcement but which seems more distant than the first one. We think it could correspond to an echo of the source present in subband 8 embedded in a deep reverberation. 

\pa It would then be quite interesting to record a single source in a room whose acoustics would be relatively quite known in order to verify if we can separate some of the reflections using the subband analysis. 

\begin{figure}[tb!]
\begin{center}
\includegraphics[width=7.5cm]{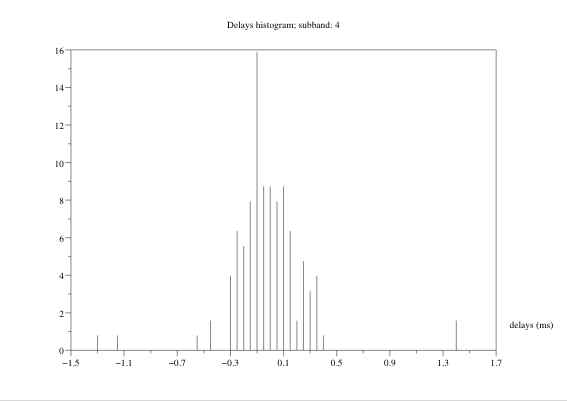}
\includegraphics[width=7.5cm]{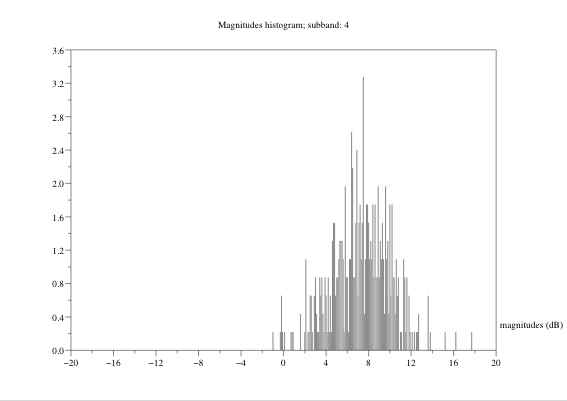} 
\caption[14]{Subband histograms for the subband 4 (400-800 Hz) in the case of the rail station real scene: relative interchannel delays (left) and attenuations (right).}
\end{center}
\end{figure} 

\pa While considering the subband 2 (see figure 15), another source points out but not as clear as for the two other subbands which make us think that this source is rather different from a point source. Listening of this subband confirms the impression as one can listen something like a reverberant rumor whose origin is quite difficult to determine: low frequency content of the announcement, part of the noise made by people present in the rail station or noise of the trains engines? A deeper analysis is needed to give access to further information. 
  
\begin{figure}[tb!]
\begin{center}
\includegraphics[width=7.5cm]{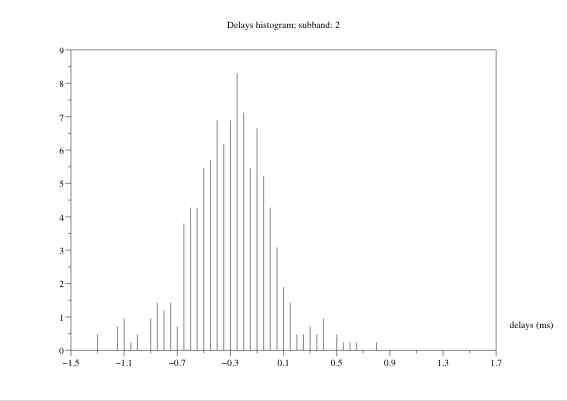}
\includegraphics[width=7.5cm]{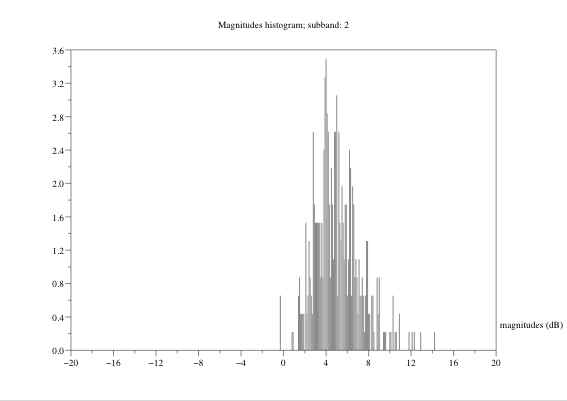} 
\caption[15]{Subband histograms for the subband 2 (50-200 Hz) in the case of the rail station real scene: relative interchannel delays (left) and attenuations (right).}
\end{center}
\end{figure} 

\pa The histograms of the other subbands are less clear but still seem to point out the presence of the same three sources. 

\pa So, even for such a rich audio scene, we seem able to identify the main sources.

\section{Discussion and perspectives}
\pa Through this paper we have presented a cartography process which may help to determine the number and the location of main sound sources.

\pa This process directly relies on a sound analysis built over a subband analysis which permits listening of any part and any approximation of the original stereophonic sound. Working with subbands would require more simple signals models, to synthesize sounds, than working with the original sound. But we think that physics-based models would be the better option to approximate sound if available. Downmixing or upmixing the audio scene, extracting sources or changing the location of the sources inside the audio scene would be effective with such physical models so their research should be a priority and we have been working on this subject.

\pa The cartography process seems quite efficient for the case of static sources as we are able to get information about at least four sources from a stereophonic recording. From our tests, we can conclude that this efficiency is rather confirmed even for recordings in live and rich ambiences (street, rail station  or market scenes for instance). But in some circumstances, further information is needed and it could be interesting to introduce a powerful parametric analysis to enhance the description of the audio scenes by tracking the little hidden sources. Using physical models should be also the right solution.

\pa For moving sources, the delays and attenuations laws derived can not be used directly to spatialize another source. Another post-processing is needed and we think that the constitution of a library of interesting motions and a neural network identification of motion based on this library can constitute a good enhancement of the cartography process. 

\pa With the cartography we are able to define the interchannel delay and attenuation for each founded source but, for the moment, we are not able to calculate the distance and the incidence direction of these sources. To derive these location, we have been working on the link between (delay, attenuation) couples and some new mathematic models of microphones. We must precise that some real measurements and recordings would be performed to test the new models for microphones. 

\pa We also work on the design of a new stereophonic recordings simulator and intend to realize a port in C or Objective-C as soon as possible. We also intend to extend this simulator to the case of a multi microphone recording set-up which may constitute a tool to understand and study sound recording for sound engineering students or to help sound engineers to build some new recording systems.

\pa We thank again all the students involved in this research by the way of their master thesis: Alexandra Carr-Brown, Maximilien Colcy, Nicolas Delatte and Antoine Valette.



\end{document}